# Duality and spatial inhomogeneity

R. Piasecki[a,*], A. Plastino[b]

[a]*Institute of Chemistry, University of Opole, Oleska 48, PL 45052 Opole, Poland*
[b]*Institute of Physics, Argentina National Research Council (CONICET) and National University La Plata, C.C. 727 (1900) La Plata, Argentina*

**Abstract**

Within the framework on non-extensive thermostatistics we revisit the recently advanced $q$-duality concept. We focus our attention here on a modified $q$-entropic measure of the spatial inhomogeneity for binary patterns. At a fixed length scale this measure exhibits a generalised duality that links appropriate pairs of $q$ and $q'$ values. The simplest $q \leftrightarrow q'$ invariant function, without any free parameters, is deduced here. Within an adequate interval $q < q_0 < q'$, in which the function reaches its maximum value at $q_0$, this invariant function accurately approximates the investigated $q$-measure, nitidly evidencing the duality phenomenon. In the close vicinity of $q_0$, the approximate meaningful relation $q + q' \cong 2q_0$ holds.

## 1. Introduction

In this paper we are concerned with the $q$-analogue of an entropic measure of spatial inhomogeneity [1,2]. The two notions of (i) spatial inhomogeneity, and (ii) complexity, are widely employed in quantitatively (qualitatively) describing morphological features of material systems. The quantity of interest is $(S_{q,\max} - S_q)$-per cell [1]. Here $S_{q,\max}$ is the highest possible value for the $q$-analogue of Shannon's celebrated logarithmic information measure. The Tsallis' index $q$ is a non-extensivity parameter characteristic of non-extensive thermostatistics [3]. Studying, at different length scales, the inhomogeneity measure with regards to computer generated binary patterns (deterministic

* E-mail address: piaser@uni.opole.pl

Sierpinski carpet (DSC) and the corresponding random patterns), initial observations [1] revealed a similarity between results corresponding to the pair of $q$ and $q' = 1/q$ values. Here we wish to give an analytical description of the non-monotonous behaviour of the $q$-measure at fixed length scales. In the neighbourhood of $q = 1$, a connection will be made with the extended $q$-duality of Naudts [4,5].

## 2. A $q$-modified inhomogeneity measure

In the important case of a $q$-entropic measure related to the spatial inhomogeneity of identical finite-sized objects (FSOs) we have given, in a previous effort [1], a full description of both (i) the micro-canonical formalism, and (ii) the pertinent averaging procedure. A brief outline follows. Consider a mixture of size-identical ($1 \times 1$ in pixels) black and white objects (particles). These particles "interact" with each other through mutual exclusion only. For a given $L \times L$ grid, let us have $n$ black particles ($0 < n < L^2$) and $m = L^2 - n$ white particles, distributed in square, non-overlapping and distinguishable $\chi = (L/k)^2$ lattice cells of size $k \times k$ (no pores). For each length scale $k$ we assume standard constraints for the cell occupation numbers, i.e., $n_1 + n_2 + \cdots + n_{\chi(k)} = n$, and, correspondingly, $m_1 + m_2 + \cdots + m_{\chi(k)} = m$, with $m_i(k) + n_i(k) = k^2$ for $i = 1, 2, \ldots, \chi(k)$. Simplifying the notation used in Ref. [1] for the so-called **B**-approach, we write $S_{\Delta,q}(k) \equiv S_{\Delta,q}(k; \mathbf{B})$. The main concomitant formulae read

$$S_{\Delta,q}(k) \equiv \frac{S_{q,\max} - S_q}{\chi} = \frac{k^2}{\chi}\left[(\chi - r_0)S_{0,q} + r_0 S_{1,q} - \sum_{i=1}^{\chi} S_{i,q}(\gamma)\right], \qquad (1)$$

where $r_0 = n \bmod \chi$, $r_0 \in \{0, 1, \ldots, \chi - 1\}$, $n_0 = (n - r_0)/\chi$, $n_0 \in \{0, 1, \ldots, k^2 - 1\}$, $\varphi_0 = n_0/k^2$, $\varphi_1 = (n_0 + 1)/k^2$, $\gamma_i(k) = n_i/k^2$. The quantities $S_{i,q}(\gamma), S_{0,q}$ and $S_{1,q}$ denote the "local" $q$-entropies written in the Tsallis' form

$$S_{i,q}(\gamma) \equiv \frac{1 - [\gamma_i^q + (1 - \gamma_i)^q]}{q - 1}, \qquad (2a)$$

$$S_{0,q} \equiv \frac{1 - [\varphi_0^q + (1 - \varphi_0)^q]}{q - 1}, \qquad (2b)$$

$$S_{1,q} \equiv \frac{1 - [\varphi_1^q + (1 - \varphi_1)^q]}{q - 1}. \qquad (2c)$$

For every length scale $1 < k < L$, there is *a special set* of the "most spatially ordered" particle distributions containing only configurations distinguished by the condition $|n_i(k) - n_j(k)| \leqslant 1$ (holding for each pair $i \neq j$). Every configuration belonging to such a *special set* represents a *reference configurational macrostate* (RCM) $\{n_i(k)\}_{\mathrm{RCM}}$, having the highest possible $S_{q,\max}(k)$-value for the configurational $q$-entropy. For the two "extreme" instances $k = (1, L)$ we have, always, $S_{q,\max}(1) = S_q(1) = 0$, and also $S_{q,\max}(L) = S_q(L) \neq 0$. Relation (1) quantifies the average deviation of a general macrostate (related to the actual system's configuration) from the reference macrostate (related to the most spatially uniform distribution). In this way we obtain a $q$-modified measure (per cell) of the spatial inhomogeneity of our mixture. For $q \to 1$ this measure reduces to the corresponding Shannon form $S_\Delta$ [1].

## 3. Extended $q$-duality transformation

Let us consider the extended inversion transformation

$$q \leftrightarrow q' \equiv \frac{q_0^\mu}{q}, \tag{3}$$

where $q_0 > 0$ and $\mu > 0$ stand, respectively, for the peak position and power parameter. For $q_0 = 1$ we have the standard "dual" transformation: $q \leftrightarrow 1/q$. Then the (non-trivial) simplest invariant function $f(q) \leftrightarrow f(q')$ must contain a sum of $q + 1/q$ like, for instance, the harmonic average of $q$ and $1/q$. If we require that the maximum value of $f(q)$ be reached at $q_0$ (the reason for this and for our use of constraints will presently become clear), necessarily we have $\mu = 2$. For a general case of $q_0 > 0$ we also require the correct limiting behaviour, that is, $f(q) \to 0$ for $q \to 0^+$ and for $q \to \infty$. Then the following form of the invariant function can be deduced

$$f(q) = \frac{\text{const}}{q + q_0^2/q}. \tag{4}$$

Now, the constant factor can be easily established if we set (c.f. Eqs. (1)–(4))

$$S_{\Delta,q}(k = \text{const}) \cong f(q), \tag{5}$$

which provides a surprisingly good approximation. The character of the approximation is quantitative in the vicinity of $q_0$ and qualitative in the physically important region $q > 0$. This approximation holds for all the model-patterns considered here. According to our previous observations [1], the maximum value of $S_{\Delta,q}(k = \text{const}) \cong f(q_0)$ equals to the corresponding Shannon measure $S_\Delta(k = \text{const})$ with a high accuracy (as we will see in Fig. 2 below). We focus attention on this property for an interval $q < q_0 < q'$. The simplest $q$-invariant form that our measure can adopt reads

$$S_{\Delta,q}(k = \text{const}) \cong S_\Delta \frac{2q_0}{q + q_0^2/q}, \tag{6}$$

where no free parameters are involved. The ratio in Eq. (6) is just an inverse-harmonic average of $q$ and $q'$. For a given binary pattern and a chosen length scale $k = \text{const}$, the "experimental" curve, i.e., the left-hand side of Eq. (6), is numerically computed according to Eq. (1). Using (i) just one point $(q_0, S_{\Delta,q=q_0}) \cong (q_0, S_\Delta)$ of this curve, (ii) an appropriate boundary conditions for the $q$-entropic measure, and (iii) its $q$-duality property, we are able to reproduce, on the basis of the right-hand side of Eq. (6), the whole experimental curve with a high degree of accuracy.

Let us now concentrate on the behaviour of $S_{\Delta,q}(k = \text{const})$ (or, equivalently, of $f(q)$), under the transformation $q \to q_0$. We can use a Taylor expansion in $q - q_0$,

up to order two, and obtain

$$S_{\Delta,q}(k=\text{const}) \cong S_\Delta - \frac{S_\Delta}{2q_0^2}(q-q_0)^2 \,. \tag{7}$$

The even parity of this function entails

$$q_0 - q + \Delta_1 = q' - q_0 - \Delta_2 \,, \tag{8}$$

where $0 < \Delta_1 < \Delta_2 \ll |q_0 - q|$ in the close vicinity of $q_0$. Thus,

$$q + q' = 2q_0 + \delta \,, \tag{9}$$

where $0 < \delta \equiv \Delta_1 + \Delta_2$ is a small correction term. In the close vicinity of $q_0$ the quite significant, approximate relation $q + q' \cong 2q_0$ holds.

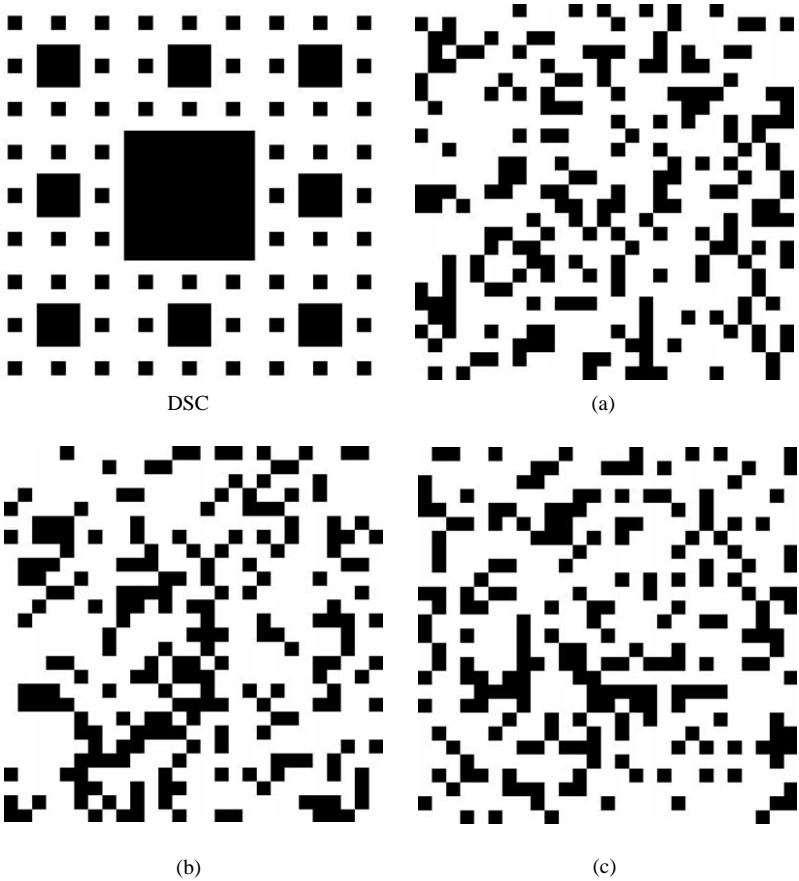

Fig. 1. The inverted DSC of size $27 \times 27$, in pixels, with $n = 729 - 512 = 217$ black particles and the corresponding pseudo-random patterns with different seeds: (a) 19520723, (b) 19781008, and (c) 19490327.

## 4. Numerical examples

Here, we will consider a few "inverted" (with respect to those considered in Ref. [1]) patterns, in which the replacement black particle ↔ white particle is effected. These simple examples clearly illustrate both the duality properties and the sensitivity of the inhomogeneity-measure to different kinds of spatial disorder. From the inverted DSC we create three pseudo-random patterns, for different seeds (a), (b) and (c) (see Fig. 1). At first sight, the spatial inhomogeneity of these patterns is very similar at every scale. However, we show in Fig. 2 that $S_{\Delta,q}(k=15)$ (solid lines) clearly distinguishes among the cases (a)–(c). Notice that the most inhomogeneous pattern, at the length scale considered here, is the (c) one, while the (a) pattern exhibits the weakest inhomogenity-degree (of about twice at the peak). The fitting by means of the curves of Eq. (6) (long dashed lines) can be regarded as a quite reasonable approximation in the whole $q$-range. The approximation given by the expansion (7) is also shown (short dashed lines), in a close vicinity of $q_0$.

For the point-object PO-measure [1] which is not considered here the similar behaviour can be observed but by means of more extended invariant function $f(q)$. This measure distinguishes between the binary initial and inverted patterns. However, it keeps the same values of $q_0$ for the corresponding patterns. The results related to the $q$-duality for both kind of patterns can be obtained around the one characteristic value of $q_0 \cong 0.15$, while for FSO-measure for the same patterns $q_0 \cong 1$.

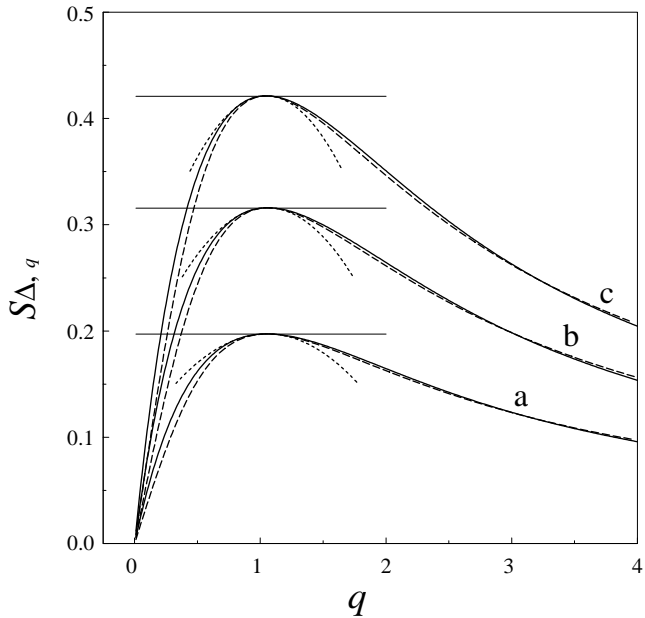

Fig. 2. Illustrative examples of the $q$-dependence of the FSO-measure (solid lines) and of the "fitting" function $f(q)$ (long-dashed lines). The length scale is $k=15$. We use the (a), (b), and (c) pseudo-random patterns depicted in Fig. 1. The parabolic approximation (short-dashed lines) works adequately only in the close vicinity of $q_0$. The corresponding Shannon's values (solid horizontal lines) are also drawn.

## 5. Conclusions

The preliminary results reported here clearly reveals the existence of a duality phenomenon for non-monotonic $q$-measures. The simple $q$-invariant function that we have deduced (whose final form is described by Eq. (6)) does not involve any free parameter. If $q_0 = 1$, then, for a sufficiently small interval $q < 1 < q'$ (the $q$-values cannot be expected to appreciably differ from unity [6]), the $\delta$-term of Eq. (9) can be neglected, and the approximate relation $q + q' \cong 2$ can be linked to a locally fluctuating cell-occupation number, provided that this fluctuating parameter is $\chi^2$-distributed. The latter assumption is necessary so as to link the deviation $(q-1)$ to the relative variance of a locally fluctuating parameter, like, for example, the temperature (see Refs. [7–9] for $q > 1$, and Ref. [10] for $q < 1$). This point deserves further research that will be reported elsewhere.

**References**

- [1] R. Piasecki, M.T. Martin, A. Plastino, Physica A (2002), in press and citations therein; cond-mat/0107471.
- [2] R. Piasecki, Physica A 277 (2000) 157;
  R. Piasecki, Surf. Sci. 454–456 (2000) 1058.
- [3] C. Tsallis, J. Stat. Phys. 52 (1988) 479;
  C. Tsallis, Physica A 221 (1995) 277;
  C. Tsallis, Braz. J. Phys. 29 (1999) 1.
- [4] J. Naudts, Chaos, Solitons, Fractals 13 (2002) 445.
- [5] L.P. Chimento, F. Pennini, A. Plastino, Phys. Rev. E 62 (2000) 7462.
- [6] R. Di Sisto, S. Martinez, F. Pennini, A. Plastino, H. Vucetich, Phys. Lett. A (2002), in press; cond-mat/0105355.
- [7] G. Wilk, Z. Włodarczyk, Phys. Rev. Lett. 84 (2000) 2770.
- [8] C. Beck, Phys. Lett. A 287 (2001) 240.
- [9] C. Beck, Europhys. Lett. (2001), in press; cond-mat/0105371.
- [10] G. Wilk, Z. Włodarczyk, Chaos, Solitons, & Fractals 13 (2002) 581.